\documentclass[12pt]{article}
\usepackage[utf8]{inputenc}
\usepackage{amsmath,amsfonts,epsfig,fp,graphicx,color}
\def\be{\begin{equation}\begin{gathered}}
\def\ee{\end{gathered}\end{equation}}
\def\Sum{\sum\limits}
\def\Prod{\prod\limits}
\def\pd{\partial}
\def\rw{\mathrm{w}}
\def\tr{\mathrm{\,tr\,}}
\def\bs{\boldsymbol}
\def\mf{\mathfrak}
\def\mc{\mathcal}

\author{P. Gavrylenko}
\title{\textbf{ Isomonodromic $\tau$-functions and $W_N$ conformal blocks}}
\date{\textit{National Research University Higher School of Economics, International Laboratory of Representation
Theory and Mathematical Physics, Moscow, Russia\\\vspace{0.4cm}
Bogolyubov Institute for Theoretical Physics, Kyiv, Ukraine\\\vspace{0.4cm}
e-mail: gavrylenko@bitp.kiev.ua}}

\begin{document}

\maketitle

\begin{abstract}
\noindent
We study the solution of the Schlesinger system for the 4-point $\mf{sl}_N$ isomonodromy problem
and conjecture an expression for the isomonodromic $\tau$-function in terms of 2d conformal field theory  beyond
the known $N=2$ Painlevé VI case. We show that this relation can be used as an
alternative definition of conformal blocks for the $W_N$ algebra and argue that the infinite number of arbitrary constants
arising in the algebraic construction of $W_N$ conformal block can be expressed in terms of only a finite set of parameters of
the monodromy data of rank $N$ Fuchsian system with three regular singular points.
We check this definition explicitly for the known conformal
blocks of the $W_3$ algebra and demonstrate its consistency with the conjectured form of the structure constants.
\end{abstract}

\section{Introduction}

There are two topics in the mathematical physics that remained independent for a long time: the theory of isomonodromic
deformations, initiated by R.~Fuchs, P.~Painlevé and L.~Schlesinger in the beginning of 20th century (see \cite{IsoReview} and references
therein), and the 2d conformal field
theory (CFT) founded by A.~Belavin, A.~Polyakov and A.~Zamolodchikov in 1984 \cite{BPZ}. Both theories have wide range of applications.
Conformal field theory describes perturbative string theory and second order phase transitions in the 2d systems. The theory of
isomonodromic deformations gives rise to
non-linear special functions such as Painlevé transcendents, which appear in different problems of mathematical physics:
for example, in the random matrix theory and general relativity.

First relations between the theory of isomonodromic deformations and 2d quantum field theory have been established in 1978-80 by M. Sato, M. Jimbo and T. Miwa \cite{SJM}. More recently, O. Gamayun, N. Iorgov and O. Lisovyy have discovered that the $\tau$-function of the 
Painlevé VI equation (related to the
rank two Fuchsian system with four regular singular points on the Riemann sphere) can be expressed as a sum of $c=1$ conformal blocks,
multiplied by certain ratios of the Barnes functions -- a typical
expansion of the correlation function in CFT \cite{GIL}.
Their formula gives the general solution of Painlevé VI equation.
This conjecture has already been proved in two ways: one proof is purely representation-theoretic and adapted initially for the
4-point $\tau$-function
\cite{ShchB} but can provide us with a collection of nontrivial bilinear relations for the $n$-point conformal blocks, whereas another one is based on the computation of monodromies of conformal blocks with degenerate fields
and allows to consider an arbitrary number of regular singular points on the Riemann sphere 
\cite{ILT}. The correspondence also extends to the irregular case: for instance, it gives exact solutions of the Painlevé V and III
equations \cite{GIL1}, \cite{irregular}, which are known to describe correlation functions in certain \textit{massive} field theories.

The present paper is concerned with the extension of the isomonodromy-CFT correspondence to higher rank. Already in \cite{GIL} there
was a suggestion that the monodromy preserving deformations of Fuchsian systems of rank $N$ should be related to 2d CFT with
central charge $c=N-1$. One obvious and natural candidate for such a theory is the Toda CFT with $W_N$ algebra of extended
conformal symmetry. We show that indeed the $N\times N$ isomonodromic problem corresponds to the $W_N$ algebra, whose Virasoro
 part has central charge $c=N-1$. These algebras were first introduced
by A. Zamolodchikov in \cite{ZamW}, and their study was substantially developed in \cite{FZ} (for the first nontrivial $W_3$-case) and \cite{FL} (for generic $W_N$).
Other developments in the theory of $W$-algebras are discussed in the review \cite{Bouwknegt}.

The most condensed form of the commutation relations of $W_3$ is given by the operator product expansions (OPEs) of the
energy-momentum tensor $T(z)$ and the $W$-current $W(z)$:

\be
\begin{split}
T(z)T(w)&=\frac{c}{2(z-w)^4}+\frac{2T\left(\frac{z+w}2\right)}{(z-w)^2}+\text{reg.}\,,\\
T(z)W(w)&=\frac{3W(w)}{(z-w)^2}+\frac{\pd W(w)}{z-w}+\text{reg.}\,,\\
W(z)W(w)&=\frac{c}{3(z-w)^6}+\frac{2T\left(\frac{z+w}2\right)}{(z-w)^4}+\\+
\frac1{(z-w)^2}\left(\frac{32}{22+5c}\right. &\left. \Lambda\left(\frac{z+w}2\right)+\frac1{20}\pd^2T\left(\frac{z+w}2\right)\right)+\text{reg.}
\end{split}
\ee
where $ \Lambda(z)=(TT)(z)-\frac3{10}\pd^2T(z)$.

The representation theory of this algebra is very similar to that of the Virasoro algebra. In the generic case one has the Verma
module with the highest vector $|\Delta,\rw\rangle$ such that $L_0|\Delta,\rw\rangle=\Delta|\Delta,\rw\rangle$, $W_0|\Delta,\rw\rangle=
\rw|\Delta,\rw\rangle$. Hence the representation space is spanned by the vectors
\be
L_{-m_1}L_{-m_2}\ldots L_{-m_k}W_{-n_1}W_{-n_2}\ldots |\Delta,\rw\rangle\label{verma},\hspace{1cm}m_1\geq m_2\geq \ldots \geq m_k, 
n_1\geq n_2\geq \ldots \geq n_k\,,
\ee
while the set of the highest weight vectors themselves corresponds to primary fields (vertex operators) of the 2d CFT.
As in the Virasoro case, these fields can be determined by their OPEs with higher-spin currents $T(z)$ and $W(z)$:

\be
\begin{split}
T(z)\phi(w)&=\frac{\Delta\phi(w)}{(z-w)^2}+\frac{\pd\phi(w)}{z-w}+\text{reg.}\\
W(z)\phi(w)&=\frac{\rw\phi(w)}{(z-w)^3}+\frac{(W_{-1}\phi)(w)}{(z-w)^2}+\frac{(W_{-2}\phi)(w)}{z-w}+\text{reg.}\label{vertex}
\end{split}
\ee

 However, the W-descendants such as $(W_{-1}\phi)$ and $(W_{-2}\phi)$ are not defined in general (this is to be
contrasted with the Virasoro case where one has e.g. $(L_{-1}\phi)(w)=\pd\phi(w)$), which means that the 3-point functions involving such fields are not really defined. As a consequence, one cannot express the matrix elements 
$$\langle\Delta_\infty,\rw_\infty|\phi(1)L_{-m_1}L_{-m_2}\ldots L_{-m_k}W_{-n_1}W_{-n_2}\ldots|\Delta_0,\rw_0\rangle$$
in terms of
$\langle\Delta_\infty,\rw_\infty|\phi(1)|\Delta_0,\rw_0\rangle$ only. It was shown
in \cite{BW3pt} in an elegant way that all such 3-point
functions can be expressed in terms of an infinite number of unknown constants
\be
C_k=\langle\Delta_\infty,\rw_\infty|\phi(1)W_{-1}^k|\Delta_0,\rw_0\rangle, \qquad k=1,2,\ldots\label{Ck}
\ee

The problem is that having this infinite number of constants (which for the 4-point conformal block actually becomes doubly infinite)
one can adjust them as
to obtain $\textit{any}$ function as a result. In this paper we show that the isomonodromic approach can fix this ambiguity
in such a way that all these parameters become functions on the moduli space of the flat connections on the sphere with 3 punctures.
In the
$\mf{sl}_3$ case this space is 2-dimensional (we denote the corresponding coordinates by $\mu$ and $\nu$), so all $C_k=C_k(\mu,\nu)$.

Note that for the $W_N$ algebra one would have the set of constants $C_{k_1,\ldots,k_l}$ with $l=\frac12(N-1)(N-2)$ non-negative indices
(e.g., this easily follows
from analysis of \cite{BW3pt}), which
is half of the dimension of the moduli space of flat $\mf{sl}_N$ connections on the 3-punctured sphere.

The paper is organized as follows. In Section 2 we briefly discuss the origins of the Schlesinger system and the space of flat
connections on the punctured Riemann sphere. Then we introduce a collection of convenient local coordinates on this space, which
are related to pants decomposition of the sphere. In Section 3 an iterative algorithm of the solution of the Schlesinger system
is proposed. We then present a set of non-trivial properties of this solution, discovered experimentally, and put forward a conjecture
about isomonodromy-CFT correspondence in higher rank, which relates $W_N$ conformal blocks to the isomonodromic tau function.
In particular, for a collection of known $W_3$ conformal blocks we present the 3-point functions that can be used to construct
the $\tau$-function in the form of explicit expansion. In Section 4 we describe the problems of definition of the general
$W_3$ conformal block and discuss how they can be addressed using the global analytic structure induced by crossing symmetry.
We conclude with a brief discussion of open questions.

\section{Isomonodromic deformations and moduli spaces of flat connections}

The main object of our study will be the Fuchsian linear system

\be
\begin{split}
\frac{d}{dz}\Phi(z)&=\Sum_{\nu=1}^{n}\frac{A_\nu}{z-z_\nu}\Phi(z)=A(z)\Phi(z)\,,\\
\Sum_\nu{A_\nu}&=0\,.\label{linearsystem}
\end{split}
\ee
Here $A_\nu$ are traceless matrices with distinct eigenvalues, $\Phi(z)$ is the matrix of $N$ independent solutions of the system normalized as $\Phi(z_0)=1$. It is obvious that upon
analytic continuation of the solutions along a contour $\gamma_\nu$ encircling $z_\nu$ they transform into some
linear combination of themselves: 

\be
\gamma_\nu: \Phi(z)\mapsto\Phi(z)M_\nu\,,
\ee
where $M_\nu\in GL_N(\mathbb{C})$. The relation $\gamma_n\ldots \gamma_1=1$ from $\pi_1(\mathbb{C}P^1\backslash\{z_1,\ldots ,z_n\},z_0)$ 
imposes the condition
\be
M_1\ldots M_n=1\,.\label{pi1}
\ee
The well-known Riemann-Hilbert problem is to find the correspondence
\be
\{M_1,\ldots ,M_n\}\rightarrow\{A_1,\ldots ,A_n\}\,.
\ee
It is easy to see that the conjugacy classes of $M_\nu$ are
\be
M_\nu\sim \exp\left({2\pi i A_\nu}\right)\,.
\ee
The eigenvalues of $A_\nu$ determine the asymptotics of the fundamental matrix solution near the singularities, so one can fix
even this asymptotics and
study the corresponding refined Riemann-Hilbert problem.
We will work only with traceless matrices $A_\nu$ since the scalar part trivially decouples.

\subsection{Schlesinger system}

Since it is difficult to solve the generic Riemann-Hilbert problem exactly, one can first ask a simpler question: how to deform simultaneously the positions
of the singularities $z_\nu$ and matrices $A_\nu$ but preserve the monodromies $M_\nu$. The
answer follows from the infinitesimal gauge transformation
\be
\begin{split}
\Phi(z)&\mapsto \left(1+\epsilon\frac{A_\nu}{z-z_\nu}\right)\Phi(z)\,,\\
A(z)&\mapsto A(z)+\epsilon\frac{A_\nu}{(z-z_\nu)^2}-\epsilon\left[\frac{A_\nu}{z-z_\nu},A(z)\right]\,,
\end{split}
\ee
that is
\be
\begin{split}
z_\nu&\mapsto z_\nu+\epsilon\,,\\
A_{\mu\neq\nu}&\mapsto A_\mu+\epsilon \frac{[A_\nu,A_\mu]}{z_\nu-z_\mu}\,,\\
A_\nu&\mapsto A_\nu-\epsilon\Sum_{\mu\neq\nu}\frac{[A_\nu,A_\mu]}{z_\nu-z_\mu}\,,
\end{split}
\ee
leading to the Schlesinger system of non-linear equations
\be
\begin{split}
\frac{\pd A_\mu}{\pd z_\nu}&=\frac{[A_\mu,A_\nu]}{z_\mu-z_\nu}\,,\\
\frac{\pd A_\nu}{\pd z_\nu}&=-\Sum_{\mu\neq\nu}\frac{[A_\mu,A_\nu]}{z_\mu-z_\nu}\,.
\end{split}
\ee
Note that one can fix $z_n=\infty$, then the corresponding matrix  $A_\infty=-\Sum_{\nu=1}^{n-1}A_\nu$
will be constant. A non-trivial statement is that the relations
\be
\frac{\pd}{\pd z_\mu}\log\tau=\Sum_{\nu\neq\mu}\frac{\tr A_\mu A_\nu}{z_\mu-z_\nu}
\ee
are compatible and define the $\tau$-function $\tau(z_1,\ldots ,z_n)$ of the Schlesinger system. 
It is easy to see that the 3-point $\tau$-function is given by a simple expression: 
 $$\tau(z_1,z_2,z_3)={\rm const}\cdot(z_1-z_2)^{\Delta_3-\Delta_1-\Delta_2}
(z_2-z_3)^{\Delta_1-\Delta_2-\Delta_3}(z_1-z_3)^{\Delta_2-\Delta_1-\Delta_3}\,,$$
where $\Delta_\nu=\frac12\tr A_\nu^2$. Let us now attempt to solve the Schlesinger system for the 4-point
case and compute the corresponding $\tau$-function in the form of certain expansion.

\subsection{Moduli spaces of flat connections}

The main object of our interest is the $\tau$-function. It depends on monodromy data which provide the full set of integrals of motion
for the Schlesinger system. It will be useful to start by introducing a convenient parametrizaton of this space.

One starts with $n$ matrices $M_\nu\in SL_N$, with fixed nondegenerate eigenvalues, i.e. there are $n(N^2-N)$ parameters. These
matrices are constrained by one equation (\ref{pi1}) and are considered up to an overall $SL_N$ conjugation, which decreases
the number of parameters by $2(N^2-1)$. So the resulting number of parameters is
\be
\dim \mathcal{M}_n^{\mf{sl}_N}(\boldsymbol\theta_1,\ldots ,\boldsymbol\theta_n)=(n-2)N^2-nN+2\,.
\ee
Here  $\boldsymbol\theta_\nu\in\mathfrak{h}$ ($\mf{h}$ is the Cartan subalgebra) define the conjugacy
classes: $M_\nu\sim e^{2\pi i\boldsymbol\theta_\nu}$.
It is obvious that $\boldsymbol\theta_\nu$ is equivalent to $\boldsymbol\theta_\nu+\boldsymbol h_\nu$, such that for all weights of the first fundamental
representation $e_i$ one has $(e_i,\boldsymbol h_\nu)\in\mathbb{Z}$. It means that $\boldsymbol h_\nu\in\oplus_{i=1}^r \mathbb{Z}
\alpha_i^\vee$,
where  $\alpha_i^\vee\in\mathfrak{h}$ are simple coroots (for the simply-laced case they coincide with the roots).

For the general Lie algebra this formula can be written as
\be
\dim \mathcal{M}_n^{\mathfrak g}(\boldsymbol\theta_1,\ldots ,\boldsymbol\theta_n)=(n-2)\dim\mathfrak{g}-n\cdot\mathrm{rank}\mathfrak{\,g}\,.
\ee
In particular, for $n=3$ punctures on the sphere
$$
\dim \mathcal{M}_3^{\mathfrak g}(\boldsymbol\theta_1,\boldsymbol\theta_2,\boldsymbol\theta_3)=
\dim\mathfrak{g}-3\cdot\mathrm{rank}\mathfrak{\,g}\,.
$$
This formula gives the number of non-simple roots of $\mathfrak{g}$. In the $\mf{sl}_N$ case it specializes to
\be
\dim \mathcal{M}_3^{\mf{sl}_N}(\boldsymbol\theta_1,\boldsymbol\theta_2,\boldsymbol\theta_3)=
(N-1)(N-2)\,.
\ee
This expression vanishes for $\mf{sl}_2$, which drastically simplifies the study of the corresponding isomonodromic problem.
However already for $\mf{sl}_3$ this dimension is equal to 2, i.e. it is nonvanishing.
One way to simplify the problem is to set $\boldsymbol\theta_2=a e_1$: in this case the orbit of the adjoint action
$$e^{2\pi i a e_1}\mapsto g^{-1}e^{2\pi i a e_1} g$$
has the dimension $\dim\mathcal{O}_{ae_1}=\dim\mathfrak{g}-\dim{\rm stab}(e_1)=N^2-1-(N-1)^2=2N-2$. The total dimension is
$2(N^2-N)+(2N-2)-2(N^2-1)=0$.   In this calculation the first two terms correspond to the dimensions of orbits: two generic and
one with a large stabilizer. The last term corresponds to one equation and one factorization. Hence
\be
\dim\mathcal{M}_3^{\mf{sl}_N}(\boldsymbol\theta_1,ae_1,\boldsymbol\theta_3)=0\,.
\ee
This case is the best known on the side of $W$-algebras \cite{Litv1}, \cite{Litv2}, \cite{Litv3}.
In the mathematical framework, this situation corresponds to rigid local systems.

\subsection{Pants decomposition of $\mathcal{M}_4^{\mathfrak{g}}$}

We begin our consideration with an arbitrary Lie group $G$ containing a Cartan torus
$H\subset G$. The corresponding Lie algebras are $\mf g$ and $\mf h$, respectively. At some point we will switch to
$G=SL_N(\mathbb{C})$ case.

The moduli space $\mathcal{M}_4^{\mathfrak{g}}$ is described by 4 matrices satisfying $M_1M_2M_3M_4=1$, defined up to conjugation:
\be
\mathcal{M}_4^{\mathfrak{g}}=\{(M_1,M_2,M_3,M_4)\}/G\,.
\ee
Let us introduce
$S=M_1M_2$ and consider two triples
\be
\{(M_1,M_2,S^{-1}),(S,M_3,M_4)\}\,.
\ee
Note that the products inside each of these triples are equal to the identity. Let us now choose the submanifold with fixed eigenvalues of
$M_1,\ldots,M_4,S$. One may also use
the freedom of the adjoint action to diagonalize $S$
$$S=e^{2\pi i\boldsymbol\sigma}\,,$$
where $\bs\sigma\in\mf{h}$. We thereby obtain a submanifold
\be
\mathcal{M}_4^{\mathfrak{g}}(\bs\theta_1,\bs\theta_2;\bs\sigma;\bs\theta_3,\bs\theta_4)=
\{(M_1,M_2,e^{-2\pi i\boldsymbol\sigma}),(e^{2\pi i\boldsymbol\sigma},M_3,M_4)\}/H\subset\mathcal{M}_4^{\mathfrak{g}}
(\bs\theta_1,\bs\theta_2,\bs\theta_3,\bs\theta_4)\,,
\ee
where the remaining factorization is performed over the Cartan torus $H\subset G$. It is very similar to
what happens for $\mathcal{M}_3^{\mathfrak{g}}$:
\be
\mathcal{M}_3^{\mathfrak{g}}=\{(M_1,M_2,M_3)\}/G=\{(M_1,M_2,e^{2\pi i\boldsymbol\theta_3})\}/H\,,
\ee
except that the  conjugation is simultaneous for both triples. To relax this condition, let us define an extra Cartan torus acting
on $\mathcal{M}_4^{\mathfrak{g}}$:
\be
h: \{(M_1,M_2,e^{-2\pi i\boldsymbol\sigma}),(e^{2\pi i\boldsymbol\sigma},M_3,M_4)\}\mapsto
\{(M_1,M_2,e^{-2\pi i\boldsymbol\sigma}),h^{-1}(e^{2\pi i\boldsymbol\sigma},M_3,M_4)h\}\,,
\ee
which looks like a relative twist of one part of the sphere with respect to another (in the $\mf{sl}_2$ case it will be exactly the
geodesic flow). Therefore one can say that
\be
\mathcal{M}^\mathfrak{g}_4(\bs\theta_1,\bs\theta_2;\bs\sigma;\bs\theta_3,\bs\theta_4)/H=\mathcal{M}^\mathfrak{g}_3(
\bs\theta_1,\bs\theta_2,-\bs\sigma)\times\mathcal{M}^\mathfrak{g}_3(\bs\sigma,\bs\theta_3,\bs\theta_4)\,.
\ee
The torus action is free, so locally it looks as a product (actually it is true even globally because the fibration
$(M_1,M_2,M_3)\mapsto(M_1,M_2,M_3)/G$
is trivial: we can give an algebraic parametrization for one representative from each conjugacy class). Therefore we have
the equality for the open subsets (denoted by $\approx$):
\be
\mathcal{M}^\mathfrak{g}_4(\bs\theta_1,\bs\theta_2;\bs\sigma;\bs\theta_3,\bs\theta_4)\approx\mathcal{M}^\mathfrak{g}_3(
\bs\theta_1,\bs\theta_2,-\bs\sigma)\times H\times\mathcal{M}^\mathfrak{g}_3(\bs\sigma,\bs\theta_3,\bs\theta_4)\,.
\ee
The above considerations suggest the following choice of coordinates on $\mathcal{M}_4^{\mathfrak{g}}$:
\begin{itemize}
\item Gluing parameters $\bs\sigma$: $\mathrm{rank\,}\mathfrak{g}$ items.
\item Invariant functions on $\mathcal{M}^\mathfrak{g}_3\times\mathcal{M}^\mathfrak{g}_3$ (for example, $\tr M_1M_2^{-1}$,
$\tr M_3^{-1}M_4$). They are invariant with respect to the action of ``relative twists'': we have
$2\dim\mathcal{M}_3^\mathfrak{g}$ such functions.
\item Relative twist parameters, which change under the twist (for example, $\tr M_2M_3^{-1}$,
$\tr M_2^{-1}M_3$), $\mathrm{rank\,}\mathfrak{g}$ items. These coordinates will be denoted by $\bs\beta\in\mathfrak{h}$.
\end{itemize}

\begin{figure}[h]
\centering\includegraphics[width=7cm]{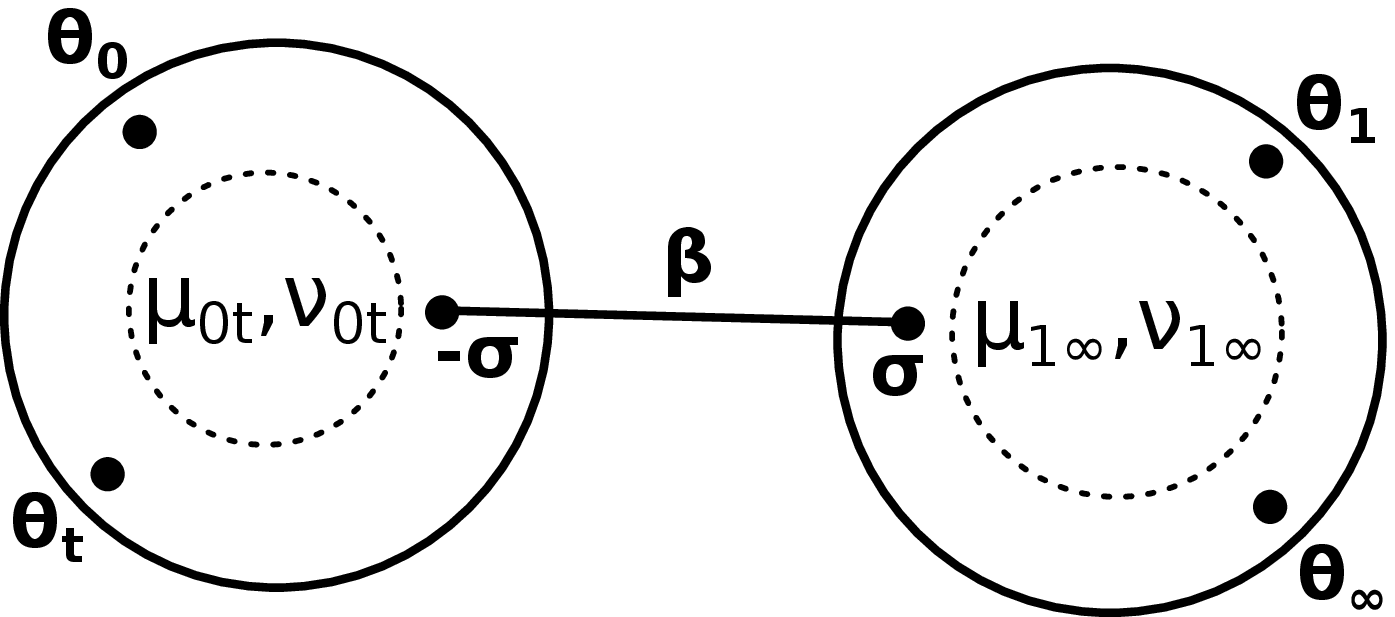}
\caption{Coordinates on $\mc{M}_4^{\mf g}$: eight = two $\bs\sigma$'s + two $\bs\beta$'s + $\mu_{0t}+\nu_{0t}+\mu_{1\infty}+\nu_{1\infty}$}
\label{fig1}
\end{figure}

This procedure is schematically depicted in Fig.\ref{fig1} for the $\mf{sl}_3$ case, where $\dim\mathcal{M}_3^{\mathfrak{sl}_3}=2$,
$\dim\mathcal{M}_4^{\mathfrak{sl}_3}=8$.  The coordinates on each copy of $\mathcal{M}_3^{\mathfrak{sl}_3}$ are denoted by $\mu,\nu$.
The indices  $\{1,2,3,4\}$ of the matrices are replaced by $\{0,t,1,\infty\}$

\subsection{Pants decomposition for $\mathcal{M}_n^\mathfrak{g}$}

Suppose that the coordinates on $\mathcal{M}_{n-1}^{\mf g}$ are chosen via the pants decomposition. Split the matrices into two
groups and define
$$S_{n-3}=M_1\ldots M_{n-2}\,,$$
\be
\begin{split}\mathcal{M}_n^\mathfrak{g}&=\{(M_1,\ldots ,M_{n-2},S_{n-3}^{-1}),(S_{n-3},M_{n-1},M_n)\}/G=\\&=
\{(M_1,\ldots ,M_{n-2},e^{-2\pi i\bs\sigma}),(e^{2\pi i\bs\sigma},M_{n-1},M_n)\}/H\approx
\mathcal{M}_{n-1}^\mathfrak{g}\times H\times\mathcal{M}_3^\mathfrak{g}\,.
\end{split}
\ee
Iteratively repeating this procedure, one is led to the following choice of coordinates on $\mathcal{M}_n^\mathfrak{g}$:

\begin{itemize}
\item $(n-3)\,\mathrm{rank\,}\mathfrak{g}$ gluing parameters $\bs\sigma_i$,
\item $(n-3)\,\mathrm{rank\,}\mathfrak{g}$ relative twist parameters $\bs\beta_i$,
\item $\Sum_{i=1}^{n-2}\dim\mathcal{M}_3^\mathfrak{g}(\bs\sigma_{i-1},\bs\theta_{i+1},-\bs\sigma_i)$ 3-point moduli of flat connections
(here we identify $\bs\sigma_0=\bs\theta_1$ and $\bs\sigma_{n-2}=-\bs\theta_n$).
\end{itemize}

Anticipating the result, let us mention that these coordinates
 are convenient from the CFT point of view: $\bs\sigma_i$ will parametrize intermediate
charges in the conformal block and $\bs\beta_i$ will be the Fourier transformation parameters. 
This description was shown to be valid in the $\mf{sl}_2$ case \cite{GIL}, \cite{ILT} and was recently demonstrated to hold for $\mf{sl}_N$
 case with $\dim\mathcal{M}_3^\mathfrak{g}=0$ \cite{unpubl}.  From a more conceptual point of view, this decomposition illustrates
that all extra parameters in the $\tau$-function expansion come from the 3-point functions.

\section{Iterative solution of the Schlesinger system}
In order to study the generic Schlesinger system, let us follow the approach proposed in the original
paper of M. Jimbo \cite{Jimbo} and in \cite[part 2]{SJM}.

Let us take the 4-point Schlesinger system and fix the singularities to be $0,t,1,\infty$. The system becomes
\be
\begin{split}
t\pd_t A_0&=[A_t,A_0]\,,\\
t\pd_t A_1&=\frac{t}{t-1}[A_t,A_1]\,,\\
\pd_t A_t&=-\frac1t[A_t,A_0]-\frac1{t-1}[A_t,A_1]\,.
\end{split}
\ee
Fixing the integral of motion $A_\infty=-A_0-A_t-A_1$, one obtains
\be
\begin{split}
t\pd_t A_0&=[A_0,A_1+A_\infty]\,,\\
t\pd_t A_1&=t(1-t)^{-1}[A_0+A_\infty,A_1]\,.\label{schles0}
\end{split}
\ee
The isomonodromic $\tau$-function is defined by
\be \pd_t\log\tau=\frac1t\tr A_tA_0+\frac1{t-1}\tr A_tA_1\,.\ee

Let us study the solution of the system (\ref{schles0}) for the case when $A_1(t)$ is finite in the limit $t\to 0$: $A_1(t)=A_1(0)+O(t^{\epsilon>0})$. Under this assumption we have
$$
t\pd_t A_0(t)=[A_0,A_\infty+A_1(0)+O(t^{\epsilon>0})]\,.
$$
If the last term were absent, then the solution would be $A_0=t^{-A_\infty-A_1(0)}\tilde A_0t^{A_\infty+A_1(0)}$. Therefore it is natural
to introduce
\be
\begin{split}
B=-A_1(0)-A_\infty&=\lim_{t\to 0}(A_0(t)+A_t(t))\,,\\
\tilde A_0(t)&=t^{-B}A_0(t)t^B\,,
\end{split}
\ee
where $\tilde A_0(t)$ has a well-defined limit as $t\to 0$. We see that in view of its definition $B$ describes
the total monodromy around $0$ and $t$
in the limit $t\to0$. Since the deformation is isomonodromic,  this monodromy is constant and
is given by $M_0M_t=M_{0t}\sim e^{2\pi iB}$. This allows to make the identification
\be
B=\bs\sigma\,.
\ee
Our system then becomes
\be
\begin{split}
t\pd_t\tilde A_0(t)&=
[\tilde A_0(t),t^{-\bs\sigma}(A_1(t)-A_1(0))t^{\bs\sigma}]\,,\\
t\pd_tA_1&=t(1-t)^{-1}[t^{\bs\sigma}\tilde A_0(t)t^{-\bs\sigma}+A_\infty,A_1(t)]\label{schles}\,.
\end{split}
\ee
Here we see an operator $t^{ad_{\bs\sigma}}$, which produces some fractional powers of $t$. It is convenient to impose the condition
that $(\bs\sigma,\bs\sigma)\ll 1$, or at least that for all roots $\alpha$  one has $|(\bs\sigma,\alpha)|<\frac12$.
This allows to organize the terms of the expansion in powers of $t$ according to their order of magnitude in the asymptotic behavior.
If necessary, one can perform an analytic continuation
of the solution from the region with small $\bs\sigma$.

We know that in the Lie algebra the operator $t^{ad_{\bs\sigma}}$ acts by
\be
\begin{split}
t^{\bs\sigma}E_\alpha t^{-\bs\sigma}&=t^{(\bs\sigma,\alpha)}E_\alpha\,,\\
t^{\bs\sigma}H_\alpha t^{-\bs\sigma}&=H_\alpha\,,
\end{split}
\ee
where $\alpha\in\mathfrak{g}^*$ is a root and $E_\alpha$,  $H_\alpha$ are the elements of the Cartan-Weyl basis. Let us
define a grading on the space of monomials $$\deg[t^{k+(\bs\sigma,\bs w)}]=(k,\bs w)\,,$$ where $\bs w\in Q_\mathfrak{g}$ is
an element of the root lattice $ Q_\mathfrak{g}=\bigoplus\limits_{i=1}^{\mathrm{rank}\mf{g}} \mathbb{Z}\alpha_i$ of $\mathfrak{g}$.
It is useful
to define a filtration 
\be
 Q^0_\mathfrak{g}\subset Q^1_\mathfrak{g}\subset Q^2_\mathfrak{g}\subset\ldots \subset Q_\mathfrak{g}
\ee
on this root lattice, which is recursively constructed as follows: $ Q^0_{\mf g}=\{0\}$, $ Q^1_{\mf g}$ is the set of all
roots of $\mf g$ and $0$, and
$$ Q^{i+1}_{\mf g}=\{\bs x+\bs y|\bs x\in Q^{i}_{\mf g},\bs y\in Q^{1}_{\mf g} \}=Q^{1}_{\mf g}+\ldots +Q^{1}_{\mf g}\,.$$
Also define the  double filtration $V_{n,m}$ on the space of all fractional-power series:
\be
t^{k+(\bs\sigma,\bs w)}\in V_{n,m}\Leftrightarrow (k\geq n)\wedge(\bs w\in Q_{\mf g}^m)\,,\\
V_{n+1,m}\subset V_{n,m}\,,\hspace{1cm}V_{n,m}\subset V_{n,m+1}\,.
\ee
Each term of the filtration is generated by these monomials. This definition turns out to be useful because of the
properties
\be
\begin{split}
t\cdot:V_{n,m}&\rightarrow V_{n+1,m}\,,\\
t^{ad_{\bs\sigma}}:V_{n,m}&\rightarrow V_{n,m+1}\,,\\
V_{n_1,m_1}\cdot V_{n_2,m_2}&\rightarrow V_{n_1+n_2,m_1+m_2}\,.
\end{split}
\ee
One can also see that the degrees present in $V_{n+1,m+k}$ are larger then in $V_{n,m}$ if $\bs\sigma$ is sufficiently small
($\forall\alpha\in Q_{\mf g}^1: |(\bs\sigma,\alpha)|<\frac1k$). We also define a slightly ambiguous notation $V_{n,\bs w}$ by
\be
t^{k+(\bs\sigma,\bs w)}\in V_{n,\bs w}\Leftrightarrow (k\geq n)\,.
\ee

Now we have all the ingredients that are necessary for the construction  of an iterative solution of the system (\ref{schles}).
Our initial data will be given by the triple of matrices $\bs\sigma$, $\tilde A_0(0)$ and $A_1(0)$.
Symbolically, the system (\ref{schles}) can be written as
\be
\tilde A_0(t)=F_0(\tilde A_0(t),A_1(t))\,,\\
A_1(t)=F_1(\tilde A_0(t),A_1(t))\,,\label{system0}
\ee
where ``affine'' bilinear (in the sense $f(x,y)=a xy+b x+c y+d$) functions $F_0$, $F_1$ have the following properties:
\be
\begin{split}
F_0&: V_{n_0,m_0}\times V_{0,0}\rightarrow0\,,\\
F_0&: V_{n_0,m_0}\times V_{n_1,m_1}\rightarrow V_{n_0+n_1,m_0+m_1+1}\subset V_{n_0+n_1,\infty}\,,\\
F_1&: V_{n_0,m_0}\times V_{n_1,m_1}\rightarrow V_{n_0+n_1+1,m_0+m_1+1}+V_{n_1+1,m_1}\subset V_{n_1+1,\infty}\label{Faction}\,.
\end{split}
\ee
Let us substitute into (\ref{system0}) the expressions
\be\tilde A_0(t)=\tilde A_0(0)+\Sum_{k=1}^\infty t^k \tilde A^k_0(t)\,,\\
A_1(t)=A_1(0)+\Sum_{k=1}^\infty t^k A_1^k(t)\,,\\
t^k \tilde A_0^k(t),t^k A_1^k(t)\in V_{k,\infty}\,.
\ee
From (\ref{Faction}) we immediately see that (\ref{schles}) takes the form
\be
\tilde A_0^k(t)=f_0^k(\tilde A_0^{<k}(t),A_1^{\leq k}(t))\,,\\
A_1^k(t)=f_1^k(\tilde A_0^{<k}(t),A_1^{<k}(t))\,.
\ee
Because of the $\leq$ sign our strategy of solving will be to compute first $A_1^k(t)$, and then subsequently determine
$\tilde A_0^k(t)$.  One can also write down explicit formulas for bilinears $f_1^k$ and $f_0^k$, which are immediate (though cumbersome)
consequences of the system (\ref{schles}).

Now let us determine which powers $(k,\bs w)$ will be actually present in the solution. This will be done again iteratively, using
only the properties (\ref{Faction}):
\begin{itemize}
\item Taking $\tilde A_0(0)\in V_{0,0}$ and $A_1(0)\in V_{0,0}$, and computing $F_1$, we get an element
of $V_{1,1}$, therefore \par $A_1\in V_{0,0}+V_{1,1}+\ldots $
\item Take $\tilde A_0(0)\in V_{0,0}$ and $A_1\in V_{0,0}+V_{1,1}+\ldots $, then $\tilde A_0\in V_{0,0}+V_{1,2}+\ldots $
\item For $\tilde A_0\in V_{0,0}+V_{1,2}+\ldots $ and $A_1\in V_{0,0}+V_{1,1}+\ldots $ one finds $A_1\in V_{0,0}+V_{1,1}+V_{2,3}+\ldots $
\item Setting $\tilde A_0\in V_{0,0}+V_{1,2}+\ldots $ and $A_1\in V_{0,0}+V_{1,1}+V_{2,3}+\ldots $ yields
$\tilde A_0\in V_{0,0}+V_{1,2}+V_{2,4}\ldots  $
\item \ldots 
\end{itemize}
Continuing this procedure one finds the following structure
\be
\begin{split}
\tilde A_0(t)&\in\Sum_{k=0}^\infty V_{k,2k}\,,\\
A_1(t)&\in V_{0,0}+\Sum_{k=1}^\infty V_{k,2k-1}\,.
\end{split}
\ee
It is easy to check that these spaces are stable under the action of $(F_0,F_1)$ described by the rules (\ref{Faction}). This is
somewhat similar to the statement that the cone is stable under the addition operation.

Indeed,
let us try to find an element of $\tilde A_0(t)$ lying in $V_{k,2k+1}$. For this one would need $n_0+n_1\leq k$, $m_0+m_1\geq 2k$, so
$m_0+m_1\geq 2(n_0+n_1)$. Since $m_1\leq 2n_1-1$ for $n_1\neq 0$ (when $F_0$ vanishes) and $m_0\leq 2n_0$, such an element cannot
exist. Similarly, for $A_1$, let us take an element lying in $V_{k,2k}$. One then needs to satisfy the constraints $n_1\leq k-1$,
$m_1\geq 2k$ (impossible) or $n_0+n_1+1\leq k$ and
$m_0+m_1+1\geq 2k$, which implies $m_0+m_1\geq 2n_0+2n_1+1$. But $m_1\leq2n_1$ and $m_0\leq2n_0$, therefore one cannot get
such an element neither.

Now let us compute the $\tau$-function and try to understand in which elements of the filtration does it lie. Since we have
\be
t\pd_t\log\tau(t)=-\tr [t^{-\bs\sigma}(A_1+A_\infty)t^{\bs\sigma}\tilde A_0+\tilde A_0^2]+t(1-t)^{-1}\tr[(A_1+A_\infty+t^{\bs\sigma}
\tilde A_0t^{-\bs\sigma})A_1]\,,
\ee
naively it could be a term in $V_{0,1}$. However, computing the constant part one finds
\be
\begin{split}
t\pd_t\log\tau(t)&=\tr(B\tilde A_0-\tilde A_0^2)+\ldots=\tr(A_tA_0)+\ldots=\\&=\frac12\tr(A_t+A_0)^2-\frac12
\tr A_0^2-\frac12\tr A_t^2+\ldots=\\
&=\frac12(\bs\sigma,\bs\sigma)-\frac12(\bs\theta_0,\bs\theta_0)-\frac12(\bs\theta_t,\bs\theta_t)+\ldots\,,
\end{split}
\ee
where $A_\nu\sim\bs\theta_\nu$. For convenience, let us introduce the notation
\be\chi=\frac12(\bs\sigma,\bs\sigma)-\frac12(\bs\theta_0,\bs\theta_0)-\frac12(\bs\theta_t,\bs\theta_t)\label{chi}\ee
The terms present in $\tr(t^{-\bs\sigma}A_1(t)t^{\bs\sigma}\tilde A_0(t))$ that are
closest to the boundary
originate from the constant part of $A_1(t)$. These terms belong to $\Sum_{k=0}^\infty V_{k,2k}$, therefore
\be\boxed{
\log\tau(t)\in\Sum_{k=0}^\infty V_{k,2k}}
\ee
Note that these estimates are too rough, since we have not taken into account that a number of the commutators actually vanish.
The actual result turns out to be the same for all three functions \par\noindent $$\log\tau,\tilde A_0,A_1\in\Sum_{k=0}^\infty V_{k,k}\,,$$
and it can be checked numerically.
Moreover, it turns out that the expansion of the $\tau$-function itself is even more restricted:
\be
\boxed{t^{-\chi}\tau(t)\in\Sum_{\bs w\in Q_\mf{g}}V_{\frac12(\bs w,\bs w),\bs w}}\label{generalsupport}\,,
\ee
which in fact provides an evidence for the 2d CFT description: different fractional powers come
from $t^\Delta$ for the different $\Delta$'s, but the conformal
dimension $\Delta=\frac12(\bs\sigma+\bs w,\bs\sigma+\bs w)$ is a quadratic function of $\bs w$ leading to the structure
(\ref{generalsupport}).

\subsection{$\mf{sl}_2$ case}

In this case we illustrate all procedures, definitions and statements using the exact solution of \cite{GIL}.

The Lie algebra $\mf{sl}_2$ is given by 3 generators $E_{\alpha},E_{-\alpha},H_{\alpha}$, such that
\be\begin{split}[E_\alpha,E_{-\alpha}]&=H_\alpha\,,\\
[H_\alpha,E_{\pm\alpha}]&=\pm2E_{\pm\alpha}\,.
\end{split}
\ee

\begin{figure}
\centering\includegraphics[width=5cm]{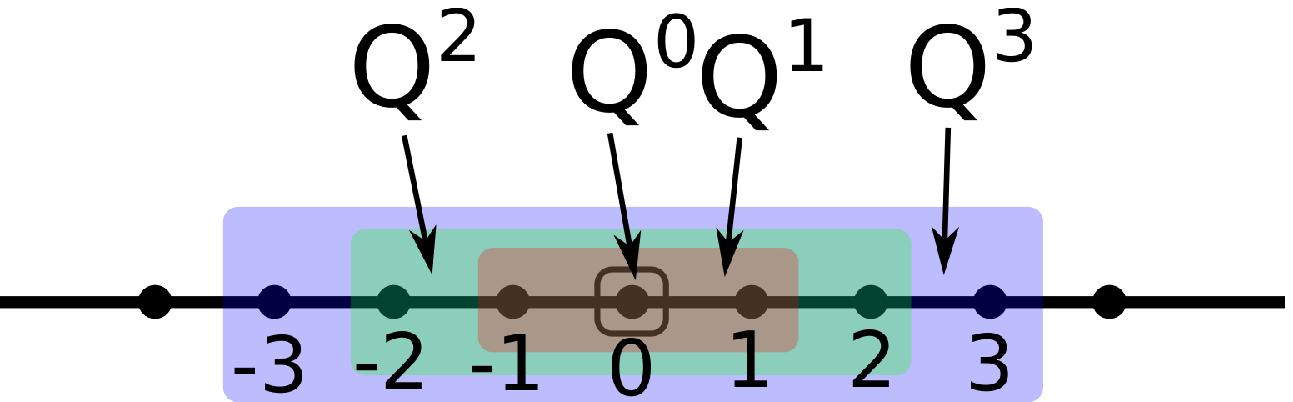}
\caption{Filtration $Q^{\bullet}_{\mf{sl}_2}$}
\label{fig2}
\end{figure}

The root lattice $Q_{\mf{sl}_2}$ is shown in Fig.\ref{fig2}.
It is spanned by one root $\alpha$. $Q_{\mf{sl}_2}^0$ is the empty
square, $Q_{\mf{sl}_2}^1$ is the red rectangle,
$Q_{\mf{sl}_2}^2$ is green and $Q_{\mf{sl}_2}^3$ is blue.

\begin{figure}[h]
\centering\includegraphics[width=7cm]{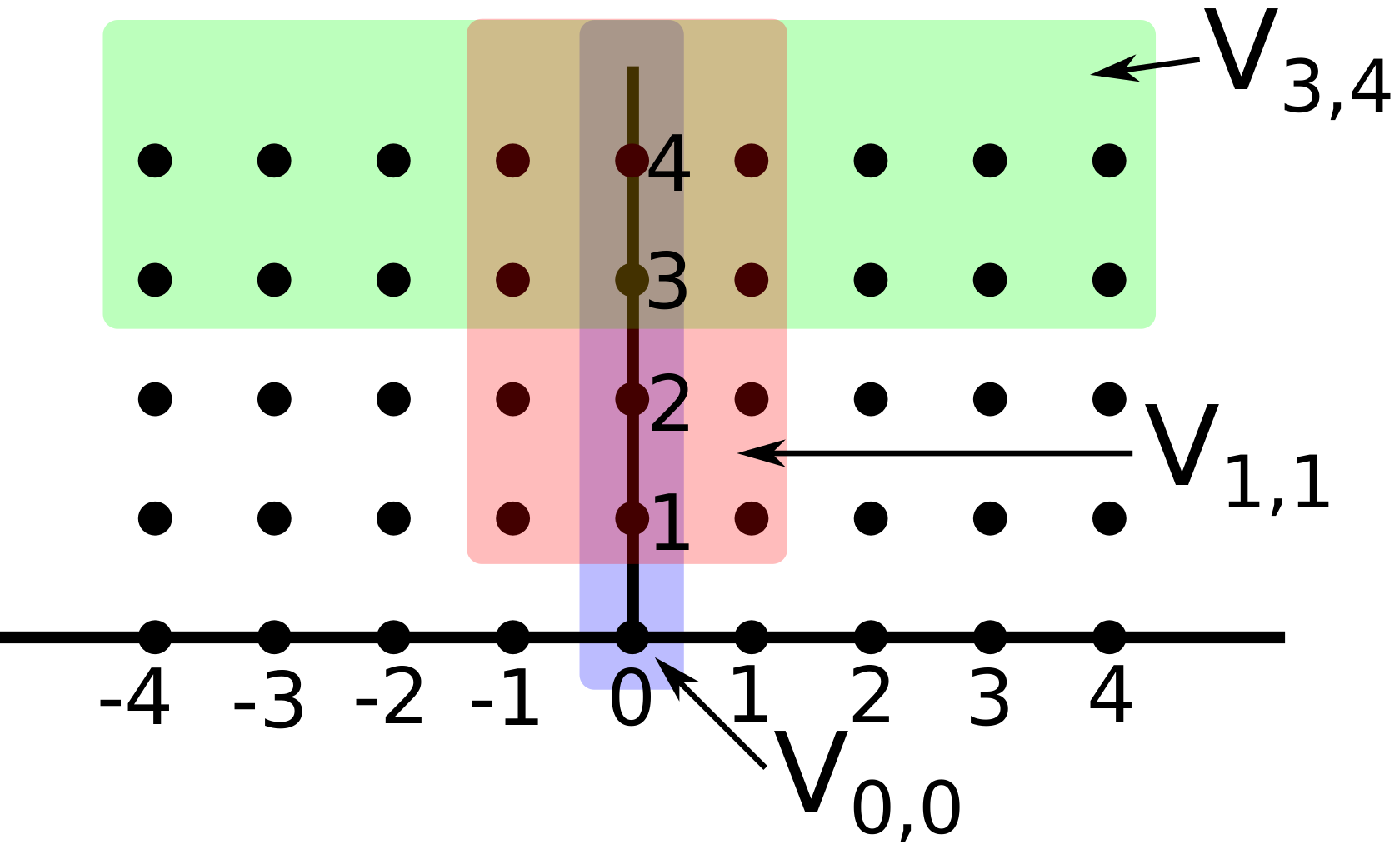}
\caption{Filtration $V_{\bullet,\bullet}$}
\label{fig3}
\end{figure}

All monomials have the form $t^{n+(\bs\sigma,\bs w)}=t^{n+m(\bs\sigma,\alpha)}$, and therefore can be depicted by the points of
a two-dimensional lattice. Note that in our normalization $(\alpha,\alpha)=2$. Several examples of the elements of this filtration
are presented in Fig.\ref{fig3}.

\noindent Here the blue region represents $V_{0,0}$, red corresponds to $V_{1,1}$ and green is $V_{3,4}$.

\begin{figure}[h!]
\centering\includegraphics[width=7cm]{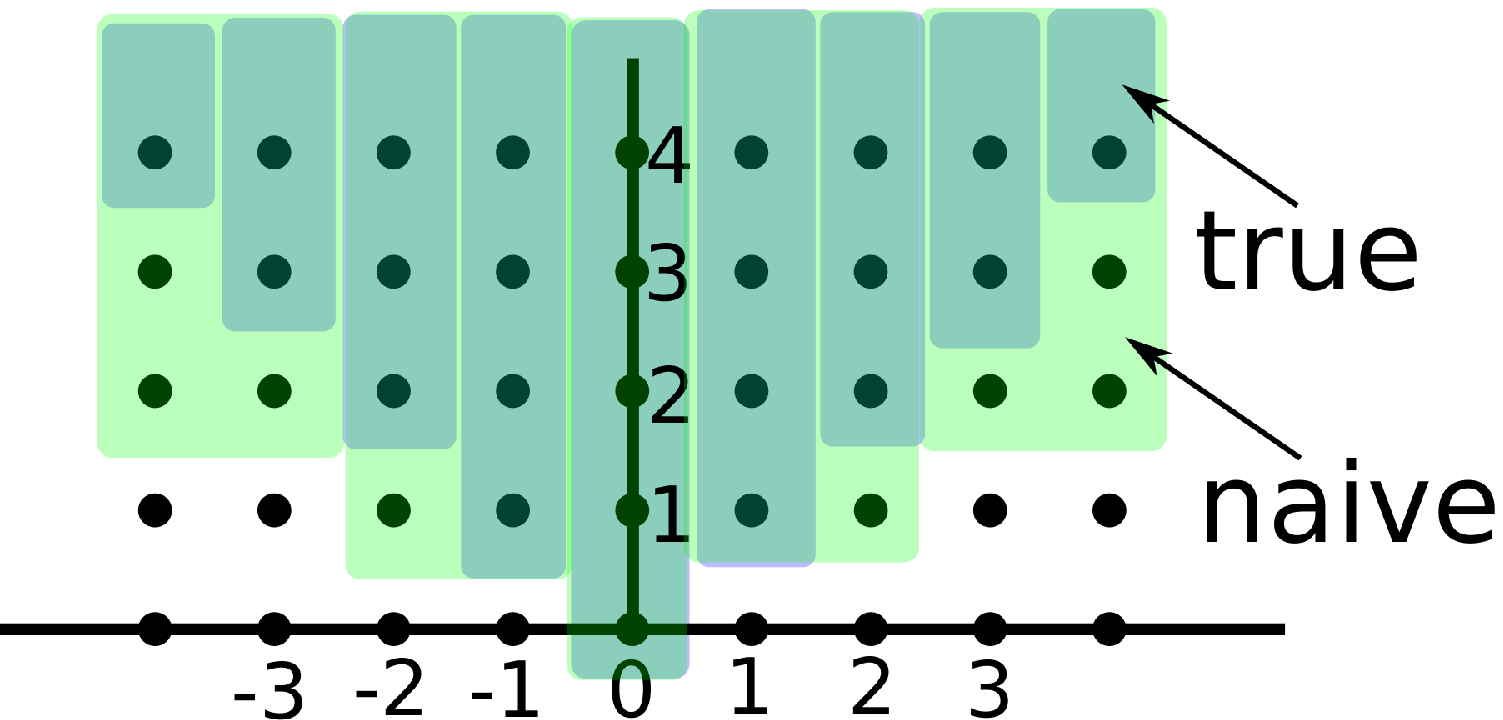}
\caption{Support of the solutions}\label{fig0}
\end{figure}
We can also show the ``true'' and ``naive'' lattice supports of the quantities $\tilde A_0(t)$, $A_1(t)$, $\log\tau(t)$ and 
$t^{-\chi}\tau(t)$. See Fig.\ref{fig0}:
green region is the ``naive'' support of $A_1(t)$, the blue region is the true support of $\tilde A_0(t), A_1(t), \log\tau(t)$, which can
be derived experimentally. Now one can use an exact formula for the tau function expansion \cite{GIL} (cf (\ref{GILtau}) below) to see that
\be
\tau(t)=t^{\sigma^2-\theta_0^2-\theta_t^2}\Sum_{k\in\mathbb Z}t^{2\sigma n}t^{n^2}f_n(t)\,,
\ee
which in turn implies
\be
t^{\theta_0^2+\theta_t^2-\sigma^2}\tau(t)\in\Sum_{k=0}^\infty V_{k^2,k}\label{mainsupport}\,.
\ee

\begin{figure}[h!]
\centering
\includegraphics[width=7cm]{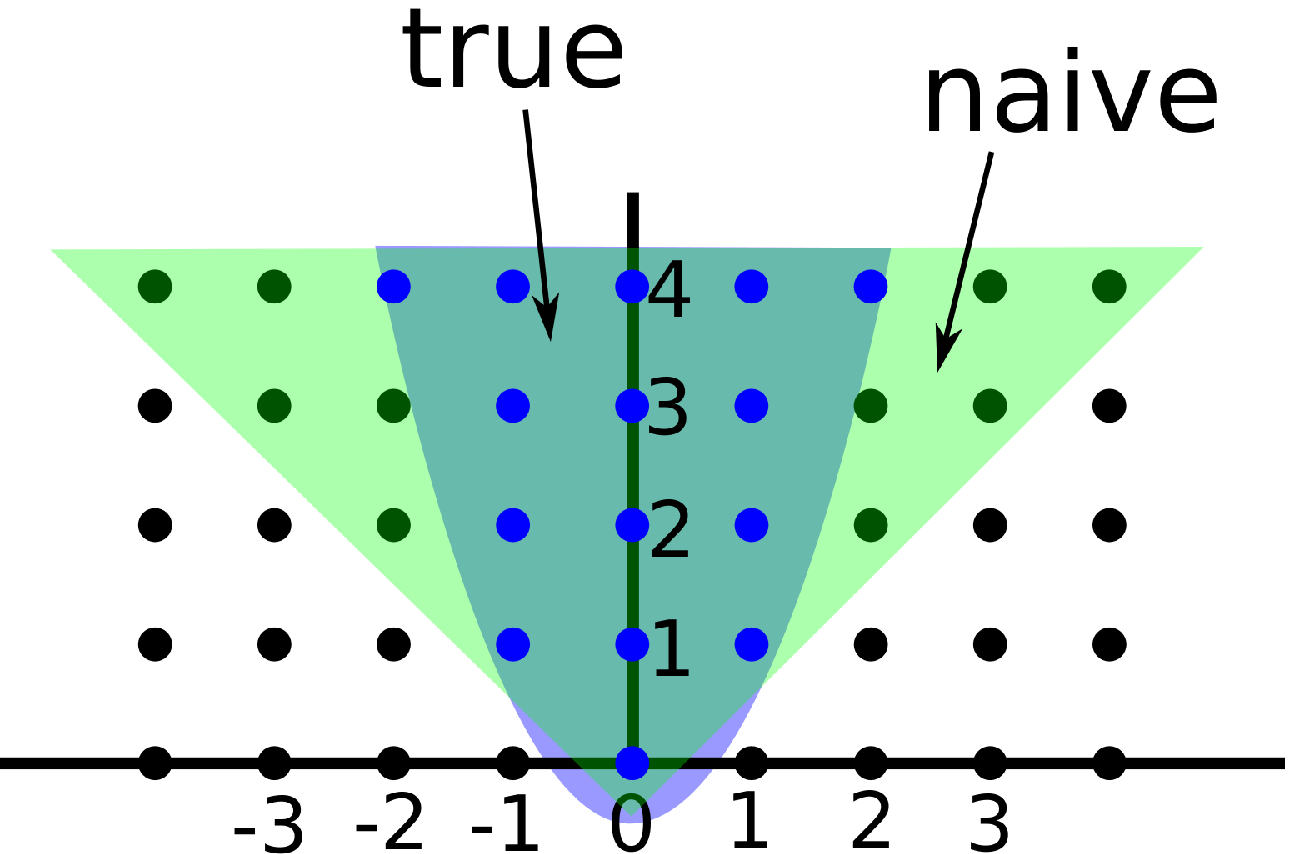}

\begin{picture}(200,60)(3,-30)
\newcommand{\weight}[4]{
\FPeval\X{(#1)*30-(#2)*15}
\FPeval\Y{(#2)*26}
\put(\X,\Y){\circle{26}
\put(-3,4){\bf #3}\put(-3,-10){#4}\put(-11,0){\line(1,0){22}}}}

\put(100,0){
\weight{0}{0}{0}{0}
\weight{1}{0}{1}{1}
\weight{-1}{0}{1}{1}
\weight{2}{0}{4}{2}
\weight{-2}{0}{4}{2}
\weight{3}{0}{9}{3}
\weight{-3}{0}{9}{3}
}

\end{picture}
\caption{Supports of $\tau(t)$ and $\log\tau(t)$: circles correspond to the integral points
of the $x$-axis, numbers inside show the $y$-coordinates of the cone and parabola.}
\label{fig4}
\end{figure}

It looks like a miracle and means that a huge number of terms cancel out when we exponentiate, but this answer confirms the
conjecture (\ref{generalsupport}). This phenomenon is illustrated in Fig.\ref{fig4} in two ways. Upper bold numbers account for the
degree in $\tau(t)$ (blue region), lower numbers correspond to the degree in $\log\tau(t)$ (green region).
Horizontal coordinate corresponds to the position in the $\mf{sl}_2$ root lattice.

Let us take the main formula from \cite{GIL}:
\be\
\tau(t)=\Sum_{n\in\mathbb{Z}} s^n C^{(0t)}_n(\theta_0,\theta_t,\sigma_{0t})
C_n^{(1\infty)}(\theta_1,\theta_\infty,\sigma_{0t})
t^{(\sigma_{0t}+n)^2-\theta_0^2-\theta_t^2}\mathcal{B}(
\{\theta_i\},\sigma_{0t}+n;t)\label{GILtau}\,,
\ee
where $\mathcal B(\ldots ;t)$ is the $c=1$ Virasoro conformal block and
\be
C^{(0t)}_n(\theta_0,\theta_t,\sigma_{0t})
C_n^{(1\infty)}(\theta_1,\theta_\infty,\sigma_{0t})=\\=\frac{\Prod_{\epsilon=\pm,\epsilon'=\pm} G(1+\theta_t+\epsilon
\theta_0+\epsilon'(\sigma_{0t}+n))G(1+\theta_1+\epsilon\theta_\infty+\epsilon'(\sigma_{0t}+n))}{G(1-2\sigma_{0t})G(1+2\sigma_{0t})}\,.
\label{GILc}\ee
Here $(\theta_\nu,-\theta_\nu)$ are the eigenvalues of the matrices $A_\nu$ in the linear system (\ref{linearsystem}),
$(e^{2\pi i\sigma_{\mu\nu}},e^{-2\pi i\sigma_{\mu\nu}})$ are the eigenvalues of $M_{\mu}M_{\nu}$, $s$ is the only variable
depending on $\sigma_{1t}$ (in a complicated way).
The main properties of (\ref{GILtau}) and (\ref{GILc}) can be summarized as follows:
\begin{enumerate}
\item The support of $\tau(t)$ is as indicated in (\ref{mainsupport}).
\item Relative twist parameter enters only via the factor $s^n$ in the structure constants.
\item The 3-point coefficients $C_n$ factorize with respect to the pants decomposition parametrization.
\end{enumerate}

We are now going to check these important properties in the $\mf{sl}_3$ case.

\subsection{$\mf{sl}_3$ case}

\begin{figure}[h!]
\centering
\includegraphics[width=7cm]{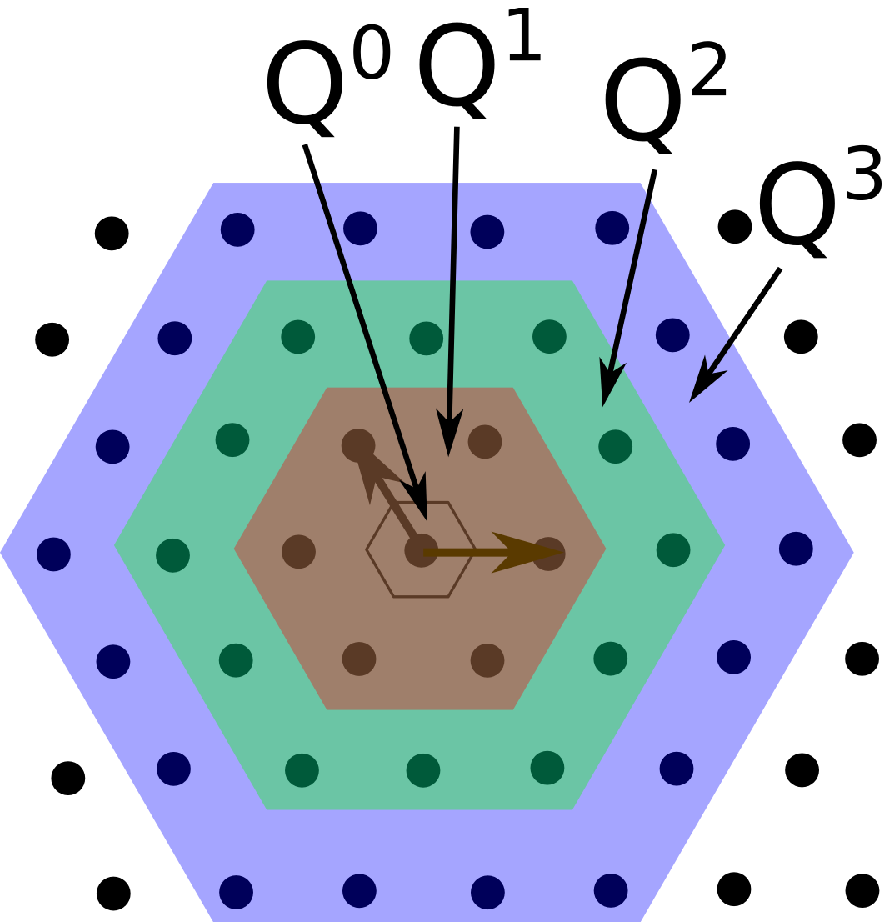}
\caption{Filtration $Q^\bullet_{\mf{sl}_3}$}
\label{fig5}
\end{figure}

Fig.\ref{fig5} illustrates the filtration on the $\mf{sl}_3$ root lattice.
The red hexagon corresponds to $Q_{\mf{sl}_3}^1$, $Q_{\mf{sl}_3}^2$ is shown in green and $Q_{\mf{sl}_3}^3$ is blue.
It is difficult to visualize  $V_{m,n}$, since one would then need a 3d picture. One can however think of 
$\Sum_{k=0}^\infty V_{k,k}$ as being a cone with hexagonal section.

Let us perform the numerical study of the $3\times3$ Schlesinger system. We  first determine which degrees $(k,\bs w)$ are
present in $\log\tau(t)$ and in $\tau(t)$ (Fig.\ref{fig7}).

\begin{figure}[h!]\centering
\begin{picture}(200,260)(-60,-80)
\newcommand{\weight}[4]{
\FPeval\X{(#1)*30-(#2)*15}
\FPeval\Y{(#2)*26}
\put(\X,\Y){\circle{26}
\put(-3,4){\bf #3}\put(-3,-10){#4}\put(-11,0){\line(1,0){22}}}}
\put(50,50){
\weight{0}{0}{0}{0}
\weight{-1}{0}{1}{1}
\weight{-1}{-1}{1}{1}
\weight{0}{1}{1}{1}
\weight{0}{-1}{1}{1}
\weight{1}{1}{1}{1}
\weight{1}{0}{1}{1}

\weight{-2}{0}{4}{2}
\weight{-2}{-1}{3}{2}
\weight{-1}{1}{3}{2}
\weight{-2}{-2}{4}{2}
\weight{0}{2}{4}{2}
\weight{-1}{-2}{3}{2}
\weight{1}{2}{3}{2}
\weight{0}{-2}{4}{2}
\weight{2}{2}{4}{2}
\weight{1}{-1}{3}{2}
\weight{2}{1}{3}{2}
\weight{2}{0}{4}{2}

\weight{-3}{0}{?}{3}
\weight{-3}{-1}{7}{3}
\weight{-2}{1}{7}{3}
\weight{-3}{-2}{7}{3}
\weight{-1}{2}{7}{3}
\weight{-3}{-3}{?}{3}
\weight{0}{3}{?}{3}
\weight{-2}{-3}{7}{3}
\weight{1}{3}{7}{3}
\weight{-1}{-3}{7}{3}
\weight{2}{3}{7}{3}
\weight{0}{-3}{?}{3}
\weight{3}{3}{?}{3}
\weight{1}{-2}{7?}{3}
\weight{3}{2}{7?}{3}
\weight{2}{-1}{7?}{3}
\weight{3}{1}{7?}{3}
\weight{3}{0}{?}{3}

\weight{-4}{0}{?}{4}
\weight{-4}{-1}{?}{4}
\weight{-3}{1}{?}{4}
\weight{-4}{-2}{?}{4}
\weight{-2}{2}{?}{4}
\weight{-4}{-3}{?}{4}
\weight{-1}{3}{?}{4}
\weight{-4}{-4}{?}{4}
\weight{0}{4}{?}{4}
\weight{-3}{-4}{?}{4}
\weight{1}{4}{?}{4}
\weight{-2}{-4}{?}{4}
\weight{2}{4}{?}{4}
\weight{-1}{-4}{?}{4}
\weight{3}{4}{?}{4}
\weight{0}{-4}{?}{4}
\weight{4}{4}{?}{4}
\weight{1}{-3}{?}{4}
\weight{4}{3}{?}{4}
\weight{2}{-2}{?}{4}
\weight{4}{2}{?}{4}
\weight{3}{-1}{?}{4}
\weight{4}{1}{?}{4}
\weight{4}{0}{?}{4}

\weight{5}{0}{?}{5}
\weight{6}{0}{?}{6}
\weight{7}{0}{?}{7}

}
\end{picture}
\caption{Degrees present in $t^{-\chi}\tau(t)$ and in $\log\tau(t)$. Number $\chi$ is given by (\ref{chi}).}
\label{fig7}
\end{figure}

As above, the upper bold numbers correspond to degrees in $t^{-\chi}\tau(t)$ and the lower ones to $\log\tau(t)$. We mark with
``?'' sign those values which are obtained at the limit of machine precision or which are greater then $7$ (so that they are not
seen in the solution up to the $7$th order). Carefully analyzing this picture, one deduces that
\be
\begin{split}
\log\tau(t)&\in\Sum_{k=0}^\infty V_{k,k}\,,\\
t^{-\chi}\tau(t)&\in\Sum_{\bs w\in Q_{\mf{sl}_3}}V_{\frac12(\bs w,\bs w),\bs w}\,.
\end{split}
\ee
It means that nonzero monomials of $\tau(t)$ fill a paraboloid, and not the naively expected cone.
 In other words, a lot of nontrivial cancellations take place, which provides further evidence for the conjecture (\ref{generalsupport}). We now list other nontrivial properties of $\tau(t)$ revealed by our experimental study.

\begin{enumerate}
\item The expansion has the form \be
\tau(t)=\Sum_{\bs w\in Q} e^{(\bs\beta,\bs w)} C^{(0t)}_{\bs w}(\bs\theta_0,\bs\theta_t,\bs\sigma_{0t},\mu_{0t},\nu_{0t})
C_{\bs w}^{(1\infty)}(\bs\theta_1,\bs\theta_\infty,\bs\sigma_{0t},\mu_{1t},\nu_{1t})\times\\\times
t^{\frac12(\bs\sigma_{0t}+\bs w,\bs\sigma_{0t}+\bs w)-\frac12(\bs\theta_0,\bs\theta_0)-\frac12(\bs\theta_t,\bs\theta_t)}\mathcal{B}_{\bs w}(
\{\bs\theta_i\},\bs\sigma_{0t},\mu_{0t},\nu_{0t},\mu_{1\infty},\nu_{1\infty};t)\,.
\ee
\item The non-zero coefficients of the expansion start from $t^{\frac12(\bs w,\bs w)}$.
\item All the dependence on the relative twist parameters is hidden in $\bs\beta\in\mf{h}$, which enters in a trivial way.
\item The dependence of structure constants on the 3-point monodromy parameters is factorized.
\item The first term in the expansion of conformal block has the form

$\mathcal{B}_0=1+[\alpha+\beta C_1(\mu_{0t},\nu_{0t})+\gamma \tilde C_1
(\mu_{1\infty},\nu_{1\infty})+\delta C_1(\mu_{0t},\nu_{0t})\tilde C_1(\mu_{1\infty},\nu_{1\infty})]t+\ldots$

This property is new, as compared to the $N=2$ case, and we will see later that it is very important.
\end{enumerate}

All these facts tell us that almost all properties of $\mf{sl}_2$ case hold in the $\mf{sl}_3$ case. This leads us to\\
\vspace{0.2cm}

\noindent\underline{\bf Main conjecture:}
$$\boxed{\mathcal{B}_{0}(\{\bs\theta_i\},\bs\sigma_{0t},\mu_{0t},\nu_{0t},\mu_{1\infty},\nu_{1\infty};t)
\text{ is a conformal block of } W_3 \text{ algebra}}$$
The corresponding dimensions and $W$-charges are given by \be\begin{split}\Delta_\nu&=\frac12(\bs\theta_\nu,\bs\theta_\nu)\,\\
\rw_\nu&=\sqrt{\frac32}\Prod_i(\bs\theta_\nu,e_i)\label{charges}\,.
\end{split}
\ee

The main advantage of the above definition of conformal block is that it depends only
on $4$ extra variables instead of a doubly-infinite set.

It is easy to check this definition for the case when $W_3$-block can be defined algebraically. This becomes
possible when $\bs\theta_t=a_t e_1$
and $\bs\theta_1=a_1 e_1$, where $e_1$ is the weight of the first fundamental representation.
The best way to present this conformal block is to use Nekrasov formulas \cite{Nekrasov} which can be applied to
conformal field theory in view of the extended AGT \cite{AGT} correspondence, first established in \cite{Wyll},
\cite{MirMor}. The most convenient (for $c=2$) expression for the conformal block can be found in \cite{Litv3}:
\be
\mathcal{B}_{\bs w}(\bs\theta_\infty,a_1,\bs\sigma,a_t,\bs\theta_0;t)=\mathcal{B}(\bs\theta_\infty,a_1,\bs\sigma+
\bs w,a_t,\bs\theta_0;t)\\
\mathcal{B}(\bs\theta_\infty,a_1,\bs\sigma,a_t,\bs\theta_0;t)=(1-t)^{\frac13 a_ta_1}\Sum_{\vec Y}t^{|\vec Y|}Z_{bif}(-\bs\theta_\infty,a_1,
\bs\sigma|\vec 0,\vec Y)\times\\\times\frac1{Z_{bif}(\bs\sigma,0,
\bs\sigma|\vec Y,\vec Y)}Z_{bif}(\bs\sigma,a_t,\bs\theta_0|\vec Y,\vec 0)\,,
\ee
where
\be
Z_{bif}(\bs\theta,a,\bs\theta'|\vec \nu,\vec\nu')=\prod_{i,j=1}^3\prod_{s\in\nu'_i}\left(
-E_{\nu'_i,\nu_j}(i(\bs\theta,e_j)-i(\bs\theta',e_i)|s)-i\frac a3\right)\times\\\times\prod_{t\in\nu_j}\left(
E_{\nu_j,\nu'_i}(i(\bs\theta',e_i)-i(\bs\theta,e_j)|t)-i\frac a3\right)\,,
\ee
and the quantities $E$ are defined by
\be
E_{\lambda,\mu}(x|s)=x-i l_\mu(s)-i a_\lambda(s)-i\,.
\ee
It yields exactly the same result as our computations using iterative solution of the Schlesinger system.


We have also conjectured in this case and checked experimentally 
a formula for the structure constants, which is a straightforward generalization of (\ref{GILc}):
\be
C^{(0t)}_{\bs w}(\bs\theta_0,a_t,\bs\sigma)C_{\bs w}^{(1\infty)}(\bs\sigma,a_1,\bs\theta_\infty)=\\
=\frac{\prod_{ij}G[1-\frac{a_t}{N}+(e_i,\bs\theta_0)-(e_j,\bs\sigma+\bs w)]G[1-\frac{a_1}{N}+(e_i,\bs\sigma+\bs w)+
( e_j,\bs\theta_\infty)]}{\Prod_i G[1+(\alpha_i,\bs\sigma+\bs w)]}\,.
\ee
Here $e_i$ denote the weights of the first fundamental representation and $\alpha_i$ are all roots of $\mf{sl}_N$ (in our case $N=3$).
This formula was recently proved \cite{unpubl} for general $N$. One can also observe a similarity between this formula and
Toda 3-point function computed in \cite{Litv1}.

\section{Remarks on $W_3$ conformal blocks}

\subsection{General conformal block}

Here we consider for simplicity the $c=2$ case, but the generalization to arbitrary $c$ is straightforward. First we explain
how the $W_N$ conformal block is defined algebraically. For that let us compute the following expression:
\be
\mathcal{B}(\bs\theta_\infty,\bs\theta_1,\bs\sigma,\bs\theta_t,\bs\theta_0;t)=
\langle-\bs\theta_\infty|\phi_{\bs\theta_1}(1)P_{\bs\sigma}\phi_{\bs\theta_t}(t)|\bs\theta_0\rangle\,,
\ee
where $|\bs\theta_0\rangle$ and $\langle-\bs\theta_\infty|$ are the highest-weight vectors  with
the charges given by (\ref{charges}), $P_{\bs\sigma}$ is the projector onto the whole Verma module (\ref{verma}) with given highest weight.
This conformal block can be computed by inserting the resolution of the identity in the Verma module. One can take, for instance, the naive basis (\ref{verma}),
or (if we do not necessarily want to preserve the $L_0$ grading) the basis from \cite{BW3pt}, or (if we wish to add the Heisenberg algebra)
the AGT basis from \cite{Litv3}, \cite{basis}.
Let us call the vectors of this basis $|\bs\sigma,\vec Y\rangle$ and suppose that $$L_0|\bs\sigma\rangle=(\Delta(\bs\sigma)+|\vec Y|)
|\bs\sigma,\vec Y\rangle\,.$$
Their scalar products will be $K_{\bs\sigma}(\vec Y,\vec Y')=\langle\bs\sigma,\vec Y|\bs\sigma,\vec Y'\rangle$. This allows to
express conformal block as
\be
\mathcal{B}(\bs\theta_\infty,\bs\theta_1,\bs\sigma,\bs\theta_t,\bs\theta_0;t)=t^\chi
\Sum_{\vec Y,\vec Y'}t^{|\vec Y|}
\langle-\bs\theta_\infty|\phi_{\bs\theta_1}(1)|\bs\sigma,\vec Y\rangle K^{-1}(\vec Y,\vec Y')\langle\bs\sigma,\vec Y'|\phi_{\bs\theta_t}(1)|\bs\theta_0\rangle
=\\=t^\chi\Sum_{\vec Y}t^{|\vec Y|}Q(\vec Y)\tilde Q(\vec Y)\,,
\ee
where $\tilde Q(\vec Y)=\Sum_{\vec Y'}K^{-1}(\vec Y,\vec Y')\langle\bs\sigma,\vec Y'|\phi_{\bs\theta_t}(1)|\bs\theta_0\rangle$ and $Q(\vec Y)=
\langle\bs\theta_\infty|\phi_{\bs\theta_1}(1)|\bs\sigma,\vec Y\rangle$ and $\chi$ is given by (\ref{chi}). The claim of
\cite{BW3pt} is that $Q(\vec Y)$ and $\tilde Q(\vec Y)$
\be
Q(\vec Y)=Q(\vec Y|C_1,\ldots ,C_{|\vec Y|})=\gamma_0(\vec Y)+\Sum_{k=1}^{|\vec Y|}\gamma_k(\vec Y) C_k\,,\\
\tilde Q(\vec Y)=Q(\vec Y|\tilde C_1,\ldots ,\tilde C_{|\vec Y|})=\tilde\gamma_0(\vec Y)+\Sum_{k=1}^{|\vec Y|}\tilde\gamma_k(\vec Y)
\tilde C_k\,,\label{triangular}
\ee
are ``triangular'' ``affine'' linear  functions of infinitely many arbitrary parameters $C_k, \tilde C_k$ defined by
\be
C_k=\langle-\bs\theta_\infty|\phi_{\bs\theta_1}(1)W_{-1}^k|\bs\sigma\rangle,\hspace{1cm}\tilde C_k=\langle\bs\sigma|W_1^k\phi_{\bs\theta_t}(1)|\bs\theta_0\rangle\,.
\ee

\subsection{Degenerate field}

Let us consider the case $\bs\theta_t=e_1$ (the weight of the first fundamental representation). The fusion rules for such fields
are known to be given by
\be
{[}e_1]\bigotimes[\bs\theta]=\bigoplus_k[\bs\theta+e_k]\label{fusion}\,.
\ee
Let us also shift $\bs\theta_0\mapsto\bs\theta_0-e_n$, multiply the conformal block by $t^{\frac{2}{3}}=t^{(e_1,e_1)}$, and define
the quantity
\be
\begin{split}
\Phi_{nk}(t)&=t^{(e_1,e_1)}\mathcal{B}(\bs\theta_\infty,\bs\theta_1,\bs\theta_0+e_k-e_n,e_1,\bs\theta_0-e_n;t)=\\&=
t^{(\bs\theta_0,e_k)+(1-\delta_{kn})}\Sum_{\vec Y}t^{|\vec Y|}Q(\vec Y,C_1,\ldots ,C_{|\vec Y|})\tilde q(\vec Y)\label{solution}\,,
\end{split}
\ee
where $\tilde q(\vec Y)$ do not contain any free parameters \cite{BW3pt}.

Now denote the degenerate field $\phi_{e_1}(t)$ by $\psi(t)$ and consider the correlation function
$$t^{(e_1,e_1)}\langle-\bs\theta_\infty|\phi_{\bs\theta_1}(1)\psi(t)|\bs\theta_0-e_k\rangle\,.$$
In the region $t\rightarrow0$ (s-channel) it can be expanded in the basis of conformal blocks written above.
But if we set $t\rightarrow1$ or
$t\rightarrow\infty$ (t- and  u-channel), then we will have the following OPEs
\be
\begin{split}
\psi(t)\phi_{\bs\theta_1}(1)&=\Sum_k C_{e_1,\bs\theta_1}^{\bs\theta_1+e_k} \cdot (t-1)^{(\bs\theta_1,e_k)}\left(\phi_{\bs\theta_1+e_k}(1)+
\text{descendants}\right)\,,\\
t^{(e_1,e_1)}\langle-\bs\theta_\infty|\psi(t)&=\Sum_k C_{\bs\theta_\infty,e_1}^{\bs\theta_\infty+e_k}\cdot
t^{-(\bs\theta_\infty,e_k)}\left(\langle-\bs\theta_\infty-e_k|+\text{descendants}\right)\,.
\end{split}
\ee
These formulas suggest that the space of conformal blocks involving $\psi(t)$ is $3$-dimensional and near each point we
have a basis with asymptotics prescribed by $\bs\theta_\nu$. It is clear that upon analytic continuation
of $\Psi_{1k}(t)$ around $0, 1, \infty$ one gets some linear combinations of the basis elements
\be
\gamma_0:\Phi_{1k}(t)\mapsto\Sum_{k'}\Phi_{1k'}(t)(M_0)_{k'k}\,,\\
\gamma_1:\Phi_{1k}(t)\mapsto\Sum_{k'}\Phi_{1k'}(t)(M_1)_{k'k}\,,\\
\gamma_\infty:\Phi_{1k}(t)\mapsto\Sum_{k'}\Phi_{1k'}(t)(M_\infty)_{k'k}\,.
\ee
In our case $M_0=\mathrm{diag}(e^{2\pi i(\bs\theta_0,e_1)},e^{2\pi i(\bs\theta_0,e_2)},e^{2\pi i(\bs\theta_0,e_3)})$.
 That these formulas \textit{must} hold can be expected on general grounds (crossing symmetry) and from the fact that the space is
3-dimensional.
However, looking at the formula (\ref{solution}), the freedom in choice of $C_k$ can give us
$\Phi_{1k}=t^{(\bs\theta_0,e_1)+(1-\delta_{k1})}\Sum_{k=0}^\infty f_lt^l$ with arbitrary $f_n$'s. It means that $W$-algebra itself 
does not account for  the global structure of conformal blocks and this information should be introduced as an extra input.

Now suppose that we have some globally-defined multivalued functions $\Phi_{1k}$. Then we have three monodromies $M_0,M_1,M_\infty$ and
one can solve the refined 3-point Riemann-Hilbert problem. Suppose that its solution is given by the matrix $F(t)$ such that
\be
\frac{d}{dt}F(t)=\left(\frac{A_0}{t}+\frac{A_1}{t-1}\right)F(t)\,,
\ee
$A_0=\mathrm{diag}\left((\bs\theta_0,e_1),(\bs\theta_0,e_2),(\bs\theta_0,e_3)\right)$ and $F(t)$ is normalized in such a way
that $F(t)=t^{A_0}(1+O(t))$. Next let us compute $R_i(t)=
\Sum_k\Phi_{1k}(t)(F(t)^{-1})_{ki}$. This
vector has the trivial monodromies around all singular points, it is regular there and $R(0)=(1,0,0)$, so that $R(t)=(1,0,0)$. It means that
\be
\boxed{\Phi_{1k}(t)=F_{1k}(t)\label{equality}}\,.
\ee

This formula allows us to fix all constants $C_k$. This is done in the following way: we solve the 3-point Riemann-Hilbert problem, take
$F_{11}(t)$ and read the coefficients of conformal block from its series. These coefficients are triangular linear combinations
(\ref{triangular}) of $C_k$ (i.e., $k$th term of the conformal block expansion involves only $C_{j\leq k}$).
This construction thus expresses $C_k$ via the moduli $(\mu,\nu)$ of flat connections on the 3-punctured sphere.
\be
\boxed{C_k=C_k(\mu,\nu)}\,.
\ee
All constants are expressed in terms of only two parameters. If we now recall the 5th experimental property of the $\tau$-function,
its origin can be easily understood: the first term of the conformal block (with the structure constants fixed above) depends only on
$C_1(\mu,\nu)$ and $\tilde C_1(\mu,\nu)$ and this dependence is at most bilinear.

\subsubsection*{Verlinde loop operators}

Here we can slightly modify our point of view: now all possible vertex operators defined by (\ref{vertex}) and (\ref{Ck})
have to be considered simultaneously. They form some $\infty$-dimensional vector space, which can be identified with
the space of 3-point conformal blocks (and which was one-dimensional in the Virasoro case). One can define the action of the Verlinde
loop operators on this space in the same way as it was done in \cite{SL3loops}. This action is given by some operators $\hat V(\gamma)$
depending on the loop $\gamma$.

If we now look at the results of \cite{ILT} then we realize that (\ref{GILtau}) can be defined alternatively as the common eigenvector of all
possible Verlinde loops. One can act in the same way for the case of 3-point conformal blocks
\be
\hat V(\gamma)\cdot\langle Y|\phi_{\bs\theta_1,\mu,\nu}(1)|Y'\rangle=M_\gamma(\mu,\nu)\cdot\langle Y|\phi_{\bs\theta_1,\mu,\nu}(1)|Y'\rangle
\ee
This procedure defines the basis of the ``right'' vertex operators $\phi_{\bs\theta_1,\mu,\nu}(1)$ characterized by some $C_k(\mu,\nu)$.
It looks more natural in this approach that $\tau$-function constructed from such operators should solve the Riemann-Hilbert problem.

The question about interpretation in $c\neq N-1$ case is still open: the problem is caused by non-commutativity of the algebra
of $\hat V(\gamma)$. Moreover, even in the minimal model-like cases $c=N-1-N(N^2-1)\frac{(k-1)^2}{k}$ for integer $k$, when the algebra
is commutative again, the relation to the isomonodromic deformations becomes unclear (see ``concluding remarks'' in \cite{ShchB} for
discussion of the Virasoro case).

\section{Conclusions}

We have discovered several important properties of the isomonodromic $\tau$-functions in higher rank, which can be interpreted as 
signatures of
the isomonodromy-CFT correspondence for the $W_N$ case. This allows to give a definition of the general $W_N$ conformal block,
depending only on a finite number of parameters.
It is also possible to prove \cite{unpubl} that the algebraic way to define well-known conformal blocks for semi-degenerate fields agrees
with the above definition.

We have also considered a particular conformal block with degenerate field and shown that its global
structure is not fixed algebraically.  The requirement of the correct global behavior of this object  yields an expression for the
whole infinite series of constants in the $W_3$ conformal block in terms of the solution of the 3-point
Riemann-Hilbert problem.

These expressions can be written in terms of coordinates on the moduli space of flat connections
on sphere with 3 punctures. This is expected to be universal
and work for any conformal block (not only for those with degenerate fields). We have checked experimentally some properties,
which support this conjecture.

Finally, let us list some remaining open problems:
\begin{itemize}
\item One needs to check that the procedure of fixing $C_k$ is self-consistent.
\item If the constants $C_k$ can be fixed in such a way, we may try to prove that the $\tau$-function can be given as a
sum of the general $W_N$ conformal blocks.
\item A constructive solution of the 3-point Riemann-Hilbert problem is still missing.
\item It would be interesting to understand the meaning of $Z_{bif}(\bs\theta_t,\bs\sigma,\bs\theta_0;\mu,\nu|\vec Y,\vec Y')$
in the context of isomonodromy-CFT correspondence. It can be
done for the case $\vec Y'=\vec 0$ and it is interesting what happens for the arbitrary Young diagrams.
\item There as an approach to the definition of conformal blocks of the light fields in the limit $c\to\infty$ \cite{Ribault}.
In that case explicit integral expression for the conformal block was derived. All the information about the 3-point functions
enters this definition via several functions of one variable. The open problem is to obtain the monodromy properties of such
conformal blocks and to identify the choice of 3-point functions that gives the conformal blocks arising in our approach.
\item It is also important to understand the meaning of the results \cite{GeomEng} about partition functions of $T_N$ theories without lagrangian description (which are believed to be the counterparts of the general $W_N$ 3-point functions) from the isomonodromic point of view.
\end{itemize}

\subsection*{Acknowledgements}

I am grateful to B. Feigin, N. Iorgov, O. Lisovyy, S. Ribault, Yu. Tykhyy for  useful discussions and comments, and especially
to A. Marshakov for the continuous support and for his interest to this work. The results of this paper were presented at the seminar
``Integrable structures in statistical and field theories'' (IITP, April 2014).

The work was supported by the joint Ukrainian-Russian (NASU-RFBR) project 01-01-14. This paper was also prepared within the framework of a subsidy granted to the National Research University Higher School of Economics by the Government of the Russian Federation for the implementation of the Global Competitiveness Program


\begin{thebibliography}{30}
\bibitem{IsoReview}A.R.~Its, V.Yu.~Novokshenov,
Lecture Notes in Mathematics {\bf 1191}, Springer-Verlag,
Berlin, (1986).


\bibitem{BPZ} A.A.~Belavin, A.M.~Polyakov, A.B.~Zamolodchikov,
Nucl. Phys. {\bf B241}, (1984), 333-380.

\bibitem{SJM} M. Sato, T. Miwa, M. Jimbo,
Publ. RIMS Kyoto Univ. {\bf 14},
(1978), 223–267; {\bf 15}, (1979), 201–278; {\bf 15}, (1979), 577–629;{\bf 15}, (1979), 871–972; {\bf 16}, (1980),
531–584.

\bibitem{GIL} O.~Gamayun, N.~Iorgov, O.~Lisovyy,
JHEP10(2012)038
[arXiv:1207.0787].

\bibitem{ShchB} M.A.~Bershtein, A.I.~Shchechkin,
[arXiv:1406.3008~[math-ph]].

\bibitem{ILT} N.~Iorgov, O.~Lisovyy, J.~Teschner,
Comm.~Math.~Phys. {\bf 336}, (2015), 671-694
[arXiv:1401.6104~[hep-th]].

\bibitem{GIL1}
O.~Gamayun, N.~Iorgov, O.~Lisovyy
J. Phys. A: Math. Theor. {\bf 46}, (2013), 335203
[arXiv:1302.1832 [hep-th]].

\bibitem{irregular}
A.~Its, O.~Lisovyy, Yu.~Tykhyy
[arXiv:1403.1235 [math-ph]].


\bibitem{ZamW}A.B.~Zamolodchikov,
Theor. Math. Phys, {\bf 65}:3, (1985), 1205–1213.



\bibitem{FZ}V.A.~Fateev, A.B.~Zamolodchikov,
Nucl. Phys. {\bf B280}, (1987), 644-660.

\bibitem{FL}V.A. Fateev, S.L. Lukyanov,
Int. J. Mod. Phys. {\bf A3} (2), (1988), 507-520.

\bibitem{Bouwknegt}P. Bouwknegt, K. Schoutens,
Phys.~Rept. {\bf 223}, (1993), 183-276,
[arXiv:hep-th/9210010].



\bibitem{BW3pt}P.~Bowcock, G.M.T.~Watts,
Theor.~Math.~Phys. {\bf 98}, (1994), 350-356
[arXiv:hep-th/9309146].

\bibitem{Litv1}V.A.~Fateev, A.V.~Litvinov,
JHEP {\bf 0711}, (2007), 002,
[arXiv:0709.3806 [hep-th]].


\bibitem{Litv2}V.A.~Fateev, A.V.~Litvinov,
JHEP {\bf 0901}, (2009), 033,
[arXiv:0810.3020 [hep-th]].

\bibitem{Litv3}V.A.~Fateev, A.V.~Litvinov,
JHEP {\bf 1201}, (2012), 051,
[arXiv:1109.4042 [hep-th]].



\bibitem{unpubl}
P.~Gavrylenko, N.~Iorgov, O.~Lisovyy,
to appear.

\bibitem{Jimbo}M.~Jimbo,
Publ. RIMS. Kyoto Univ. {\bf18} (1982), 1137-1161.




\bibitem{Nekrasov}N. Nekrasov,
Adv.~Theor.~Math.~Phys. {\bf 7}, (2004), 831-864,
[arXiv:hep-th/0206161].

\bibitem{AGT}L. Alday, D. Gaiotto, Y. Tachikawa,
Lett.~Math.~Phys.~{\bf 91}, (2010), 167-197,
[arXiv:0906.3219 [hep-th]].


\bibitem{Wyll}N.~Wyllard,
JHEP {\bf 0911}, (2009), 002,
[arXiv:0907.2189 [hep-th]].

\bibitem{MirMor}A.~Mironov, A.~Morozov,
Nucl.~Phys.~{\bf B825}, (2010), 1-37,
[arXiv:0908.2569 [hep-th]].


\bibitem{basis}A.A.~Belavin, M.A.~Bershtein, B.L.~Feigin, A.V.~Litvinov, G.M.~Tarnopolsky,
Comm.~Math.~Phys.{\bf 319}, (2013), 269-301,
[arXiv:1111.2803 [hep-th]].



\bibitem{SL3loops}I. Coman, M. Gabella, J. Teschner,
[arXiv:1505.05898 [hep-th]]


\bibitem{Ribault}V.~Fateev, S.~Ribault,
JHEP {\bf 0212}, (2012), 001,
[arXiv:1109.6764 [hep-th]].

\bibitem{GeomEng}L.~Bao, V.~Mitev, E.~Pomoni, M.~Taki, F.~Yagi
JHEP {\bf 0114}, (2014), 137,
[arXiv:1310.3841 [hep-th]]

\end{thebibliography}
\end{document}